\documentclass[journal=nalefd,manuscript=letter,layout=twocolumn]{achemso}
\usepackage[colorlinks,linkcolor={blue},citecolor={blue},urlcolor={blue}]{hyperref} 
\usepackage[T1]{fontenc}
\usepackage[utf8]{inputenc}
\usepackage{amssymb}
\usepackage{braket}
\usepackage{array}
\usepackage{multirow}
\usepackage{amsmath}
\usepackage{amsfonts}
\usepackage{graphicx}
\usepackage{xcolor}
\usepackage{bm}
\usepackage{bbm}
\usepackage{subfigure}
\usepackage{float}
\usepackage[normalem]{ulem}
\usepackage{mathtools} 

\title{Interlayer electric multipole Hall effect in twisted multilayers}
\author{Chengxin Xiao}
\affiliation{New Cornerstone Science Laboratory, Department of Physics, The University of Hong Kong, Hong Kong, China}
\alsoaffiliation{School of Electrical and Information Engineering, Zhengzhou University, Zhengzhou, Henan 450001, China}
\alsoaffiliation{HK Institute of Quantum Science \& Technology, The University of Hong Kong, Hong Kong, China}
\author{Cong Xiao}
\affiliation{Interdisciplinary Center for Theoretical Physics and Information Sciences (ICTPIS), Fudan University, Shanghai 200433, China}
\alsoaffiliation{State Key Laboratory of Surface Physics, Fudan University, Shanghai 200433, China}
\author{Dawei Zhai}
\email{dzhai@hku.hk}
\affiliation{New Cornerstone Science Laboratory, Department of Physics, The University of Hong Kong, Hong Kong, China}
\alsoaffiliation{HK Institute of Quantum Science \& Technology, The University of Hong Kong, Hong Kong, China}
\alsoaffiliation{State Key Laboratory of Optical Quantum Materials, The University of Hong Kong, Hong Kong, China}
\author{Wang Yao}
\email{wangyao@hku.hk}
\affiliation{New Cornerstone Science Laboratory, Department of Physics, The University of Hong Kong, Hong Kong, China}
\alsoaffiliation{HK Institute of Quantum Science \& Technology, The University of Hong Kong, Hong Kong, China}
\alsoaffiliation{State Key Laboratory of Optical Quantum Materials, The University of Hong Kong, Hong Kong, China}

\begin{document}

\begin{abstract}
{Electrons in layered van der Waals materials possess a layer pseudospin characterizing their wave-function distribution among layers. In twisted structures, this pseudospin forms nontrivial textures, leading to intriguing phenomena such as the layer Hall effect (LHE), where distinct layer Hall currents flow despite the presence of time-reversal symmetry. In chiral bilayers, LHE manifests as an interlayer electric dipole Hall effect with Hall counterflows and a concomitant in-plane magnetic dipole. Multilayers host richer layer-dependent Hall currents, generating interlayer electric multipole Hall effects and in-plane magnetic multipoles. We start from exploring the interlayer electric quadrupole Hall effect in mirror-symmetric twisted trilayers. At small twist angles, interlayer translation efficiently tunes layer Hall current magnitudes. At large angles and low doping, the currents can be well accounted for by adding the contributions from the two individual twisted interfaces. This decomposition allows obtaining layer-resolved Hall currents in large-angle twisted multilayers even without well-defined periodicity.}
\end{abstract}
\maketitle

\section{Introduction}
Moir\'e superlattices of 2D materials formed by lattice mismatch or misorientation have emerged as versatile platforms for exploring various exciting physics phenomena. Celebrated groundbreaking discoveries include superconductivity in magic-angle twisted bilayer graphene (TBG)~\cite{TBGCaoYuan2018a} and fractional quantum anomalous Hall effects in twisted bilayer MoTe$_2$~\cite{FCIMoTe2Jiaqi2023,FCIMoTe2ShanJie2023,FCIMoTe2Park2023,FCIMoTe2PRX2023}, which are made possible by the presence of flat energy bands boosting electron interactions.
The chiral structure of twisted materials also leads to interesting effects that do not require strong electron interaction, for example, chiral optical and transport responses~\cite{CDTBGNatNano2016,Brey2DM2017,TobiasPRL2018,TobiasPRB2018,CDTBGslidingMelePRB2019,CDTwistedhBNPRL2020,TobiasPRB2020,TobiasNanoscale2020,TobiasNanoLett2020,TBGYangBinghaiInnovation2021,ZhaiLayerHallNC2023,ZhuJihangLayerHall2024,ChenCongPRR2024,LiJuncheng2024,YaoYuguiLayerNernstPRB2024,TobiasTrilayerPRB2024,LiCiChiralExcitonHall2024,Okyay2025,Okyay2022}. It has been shown that twisted homobilayers with strong chiral interlayer coupling (e.g., TBG and twisted MoTe$_2$) can support large counterflowing linear Hall currents in the two layers in the presence of time-reversal symmetry (TRS)~\cite{TobiasPRL2018,TobiasPRB2018,TobiasPRB2020,ZhaiLayerHallNC2023,ZhuJihangLayerHall2024}, leading to the layer Hall effect (LHE) beyond magnetic systems~\cite{LayerHallNature2011,LayerHallAdvSci2021,LayerHallNSR2022,LayerHallPRB2022,LayerHallNSR2023,LayerHallFanFengrenNC2024,LayerHallYaoYuguiPRB2025,LayerHallQiaoZhenhuaPRL2025,LayHallMaYandongNC2025}. In twisted bilayers the layer Hall counterflows yield an in-plane magnetic dipole moment, underlying the emergence of circular dichroism in the quasi-2D limit.
The Hall counterflows in a twisted bilayer can also be regarded as an interlayer electric dipole Hall current~\cite{ZhaiLayerHallNC2023}. In multilayer structures, the layer-dependent Hall currents are expected to exhibit richer patterns among different layers~\cite{HuiyuanMultipole2024,TobiasTrilayerPRB2024}, leading to, e.g., interlayer electric quadrupole Hall currents and in-plane magnetic quadrupole moments in trilayers [Fig.~\ref{fig1}(a)]. Here we explore such LHE in the presence of TRS, by employing twisted trilayer graphene (TTG) as a model system. In mirror-symmetric TTG where the two twisted interfaces have opposite rotations~\cite{TTGAshvinPRB2019,TTGNanoLett2020,MATTGPabloNature2021,MATTGPhilipKimScience2021}, the LHE is manifested as a pure interlayer electric quadrupole Hall effect, as the dipole contribution is prohibited by the chiral structure. 
Interestingly, the charge Hall flows in the two outer layers have the same direction and magnitude, and the compensation by the hidden flow in the middle layer ensures that the overall charge Hall current vanishes.
When an interlayer translation breaks the out-of-plane mirror symmetry, a pure electric quadrupole Hall effect is preserved, whose magnitude can be tuned by the translation when the twist angle is small. In large-angle twisted TTG, as interlayer coupling becomes weak at low energies, we demonstrate that the layer-resolved Hall currents can be accurately obtained by adding the contributions from the two individual twisted interfaces, which have modest dependence on interlayer translation. Additionally, the layer Hall current contributed by large angle twisted interface follow a universal scaling behavior with respect to the twist angle and Fermi energy. These properties enable tailor-made layer-resolved Hall responses, or equivalently, electric multipole Hall responses, in general twisted multilayers with arbitrary and possibly incommensurate twist angles.


\section{Interlayer electric multipole Hall effect \& in-plane magnetic multipole}

For a multilayer system, the current in a specific layer $l$ can be calculated by~\cite{ZhaiLayerHallNC2023,LayerHallFanFengrenNC2024}
\begin{equation}
	\bm{j}^{l}=\frac{e}{4\pi^2}\sum_{n}\int f_n(\bm{k})\bm{v}_{n}^{l}(\bm{k})\,d\bm{k},
\end{equation}
where $e<0$ is the charge of an electron, $f_n$ is the Fermi-Dirac distribution function on the $n$-th band with energy $\varepsilon_n(\bm{k})$ and Bloch state ${\ket{\psi_{n\bm{k}}}}=e^{i\bm{k}\cdot\bm{r}}\ket{u_{n\bm{k}}}$, and $\bm{v}_n^l(\bm{k})=\braket{u_{n\bm{k}}|\hat{\bm{v}}^l|u_{n\bm{k}}}=\braket{u_{n\bm{k}}|\frac{1}{2}\{\hat{\bm{v}},P_l\}|u_{n\bm{k}}}$ is the velocity projected onto the layer $l$ with $P_l=\ket{l}\bra{l}$.
Within the Boltzmann transport theory and the constant relaxation time $\tau$ approximation, the distribution function reads $f_n\approx f_{\rm eq}-\frac{e}{\hbar}\tau \boldsymbol{E}\cdot\partial_{\bm{k}}f_{\rm eq}$ to the linear order, where $\bm{E}$ is the applied electric field and $f_{\rm eq}$ is the equilibrium Fermi-Dirac distribution.
We will focus on time-reversal symmetric systems and linear responses in the following. In such scenarios, only the band velocity $\bm{v}_n(\bm{k})=\braket{u_{n\bm{k}}|\hat{\bm{v}}|u_{n\bm{k}}}=\hbar^{-1}\braket{u_{n\bm{k}}|\partial_{\bm{k}}\hat{H}(\bm{k})|u_{n\bm{k}}}$ can contribute to a nonzero current, and the transverse conductivity in layer $l$ becomes
\begin{equation}
	\sigma_{ij}^{l}=-\frac{e^{2}}{h}\frac{\hbar\tau}{2\pi}
	\sum_{n}\int\frac{\partial f_{\rm eq}}{\partial \varepsilon_n}v_{n,i}^{l}(\bm{k})v_{n,j}(\bm{k})\,d\bm{k},
\end{equation}
where $i\ne j\in\{x,\,y\}$. Then the Hall conductivity of layer $l$, defined by $\sigma_{H}^{l}
=\frac{1}{2}(\sigma_{xy}^{l}-\sigma_{yx}^{l})$, reads
\begin{align}
        \sigma_{H}^{l}
        =\frac{e^{2}}{h}\frac{\hbar\tau}{4\pi}
        \sum_{n}\int\frac{\partial f_{\rm eq}}{\partial \varepsilon_n}\left[\bm{v}_{n}(\bm{k})\times\bm{v}_{n}^{l}(\bm{k})\right]_{z}d\bm{k},~\label{eq:LayerHallCond}
\end{align}
and the corresponding Hall current is $\bm{j}_H^l=\sigma_H^l\hat{\bm{z}}\times\bm{E}$.
In this work we will focus on trilayer systems and comment on the generalization to more complicated multilayers at the end. In the case of TTG models considered in the following, the band velocity operators have simple diagonal structures in the layer space: $\hat{\bm{v}}=\hat{\bm{v}}^t+\hat{\bm{v}}^m+\hat{\bm{v}}^b$, where $\hat{\bm{v}}^t=\text{diag}(\hat{\bm{v}}_1,0,0)$, $\hat{\bm{v}}^m=\text{diag}(0,\hat{\bm{v}}_2,0)$, $\hat{\bm{v}}^b=\text{diag}(0,0,\hat{\bm{v}}_3)$.
$t$, $m$ and $b$ labels the top, middle and bottom layer respectively, $\hat{\bm{v}}_{1\sim3}$ can be easily identified once the Hamiltonian is explicitly specified.

\begin{figure}[t]
    \centering
    \includegraphics[width=1\linewidth]{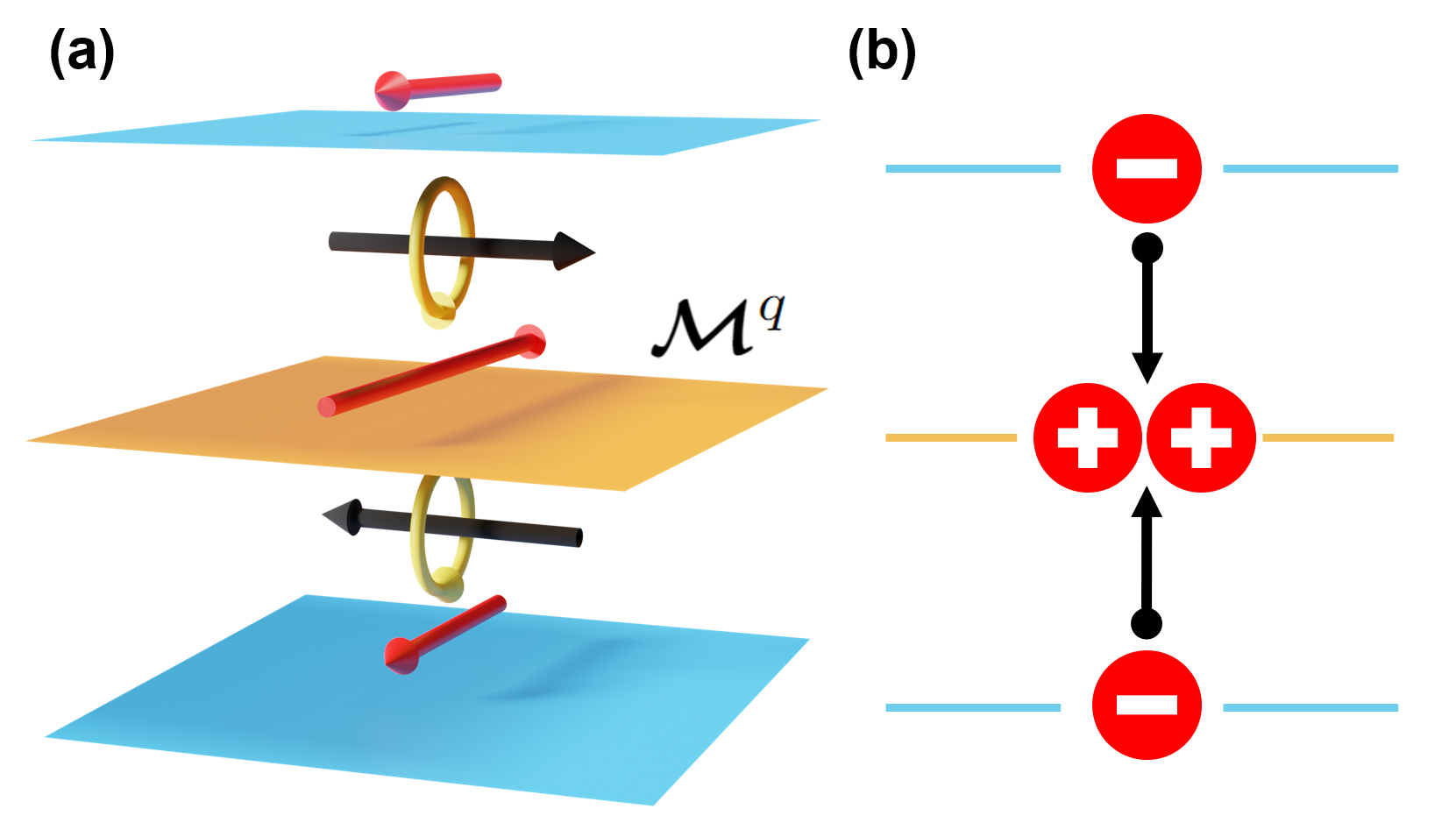}
    \caption{Schematics of layer-dependent Hall effects in a mirror-symmetric trilayer. (a) The Hall currents in the top and bottom layers are identical (shorter red arrows), while the current in the middle layer flows oppositely with doubled magnitude (longer red arrow). Such layer-dependent current flows correspond to the Hall current of interlayer electric quadrupole (b), which also generates a in-plane magnetic quadrupole moment composed of two opposite in-plane magnetic dipole moments contributed by current counterflows in adjacent layers. {The yellow ring with black arrow indicates the direction of in-plane magnetic dipole at each twisted interface.} }\label{fig1}
\end{figure}

The layer-dependent Hall currents in a multilayer can also be understood as the Hall currents of interlayer electric multipoles~\cite{HuiyuanMultipole2024}.
For trilayer systems, one can define the interlayer electric dipole and quadrupole operators as $\hat{\mathcal{E}}_\perp^d=\text{diag}(1,0,-1)$ and $\hat{\mathcal{E}}_\perp^q=\text{diag}(1,-2,1)$, the latter of which is schematically shown in Fig.~\ref{fig1}(b). The corresponding electric dipole and quadrupole Hall currents read $\bm{j}_H^{d,q}=\sigma_H^{d,q}\hat{\bm{z}}\times\bm{E}$, where 
\begin{equation}
    \sigma_{H}^{d/q}=\frac{e^{2}}{h}\frac{\hbar\tau}{4\pi}
    \sum_{n}\int\frac{\partial f_{\rm eq}}{\partial \varepsilon_n}\left[\bm{v}_{n}(\bm{k})\times\bm{v}_{n}^{d/q}(\bm{k})\right]_{z}d\bm{k}
\end{equation}
is the electric dipole/quadrupole Hall conductivity, with
$\bm{v}_n^{d/q}(\bm{k})=\braket{u_{n\bm{k}}|\hat{\bm{v}}^{d/q}|u_{n\bm{k}}}$ being the dipole/quadrupole velocity:
\begin{equation}
	\begin{aligned}
	\hat{\bm{v}}^{d}
	&=\frac{1}{2}\{\hat{\bm{v}},\hat{\mathcal{E}}_\perp^d\}
	=
	\begin{pmatrix}
	\hat{\bm{v}}_1&0&0\\
	0&0&0\\
	0&0&-\hat{\bm{v}}_3
	\end{pmatrix},\\
	\hat{\bm{v}}^{q}
	&=\frac{1}{2}\{\hat{\bm{v}},\hat{\mathcal{E}}_\perp^q\}
	=
	\begin{pmatrix}
	\hat{\bm{v}}_1&0&0\\
	0&-2\hat{\bm{v}}_2&0\\
	0&0&\hat{\bm{v}}_3
	\end{pmatrix}.
	\end{aligned}
\end{equation}
The electric dipole and quadrupole Hall conductivities are related to the layer Hall conductivities [Eq.~(\ref{eq:LayerHallCond})] by
\begin{alignat}{2}
	\sigma_{H}^{t}&+\sigma_{H}^{m}+\sigma_{H}^{b}=0~~~~&&\sigma_{H}^{t}=\frac{1}{6}\sigma_{H}^{q}+\frac{1}{2}\sigma_{H}^{d}  \nonumber \\
	\sigma_{H}^{d}&=\sigma_{H}^{t}-\sigma_{H}^{b}~~~~~~~~~~~~\Longleftrightarrow ~~~~&&\sigma_{H}^{m}=-\frac{1}{3}\sigma_{H}^{q}    \nonumber\\
	\sigma_{H}^{q}&=\sigma_{H}^{t}+\sigma_{H}^{b}-2\sigma_{H}^{m}~~~~&&\sigma_{H}^{b}=\frac{1}{6}\sigma_{H}^{q}-\frac{1}{2}\sigma_{H}^{d}   \nonumber
\end{alignat}
where $\sigma_{H}^{t}+\sigma_{H}^{m}+\sigma_{H}^{b}=0$ corresponds to a vanishing net linear charge Hall response in the presence of TRS as dictated by the Onsager reciprocal relation~\cite{Onsager}. 
One identifies that an interlayer electric dipole (quadrupole) Hall response implies $\sigma_{H}^{t}\ne\sigma_{H}^{b}$ ($\sigma_{H}^{m}\ne0$) or vice versa. 
A nonzero electric dipole Hall response requires the system to be chiral without inversion and mirror symmetries~\cite{ZhaiLayerHallNC2023}. 
Thus, twisted trilayers with out-of-plane mirror symmetry are ideal platforms for studying the pure electric quadrupole Hall effect.

Meanwhile, the presence of layer-dependent Hall currents gives rise to magnetic dipole and quadrupole moments, {which might be probed by X-ray magnetic circular dichroism~\cite{hchg-7bq9,doi:10.1126/science.abd5146}}.  
Due to the infinite extension of the current distribution in the $xy$-plane, only the in-plane magnetic dipole and the magnetic quadrupole associated with the $z$-coordinate are physically well-defined. 
This necessitates the definition of in-plane magnetic dipole and quadrupole vector densities, $\bm{\mathcal{M}}^d$ and $\bm{\mathcal{M}}^q$, by expanding the magnetic coupling energy $\mathfrak{E} = -\int \bm{j} \cdot \bm{A} \, dz$ between the in-plane current and an in-plane magnetic field with a weak $z$-gradient and extracting the dipole and quadrupole terms $\mathfrak{E}_{d} = -\bm{\mathcal{M}}^d \cdot \bm{B}$ and $\mathfrak{E}_{q} = -\frac{1}{2}\bm{\mathcal{M}}^q \cdot \partial_z \bm{B}$. Specifically,
\begin{align}
    \bm{\mathcal{M}}^d & =\sum_{l} z_l\hat{\bm{z}}\times \bm{j}_H^l, \\
    \bm{\mathcal{M}}^q & =\sum_{l} z_l^2(\hat{\bm{z}}\times \bm{j}_H^l),
\end{align}
where $z_l$ is the out-of-plane coordinate of layer $l$ and the coordinate origin is placed in the middle layer.
In the linear response regime, we get
\begin{align}
    \bm{\mathcal{M}}^d & =-d_0(\sigma_H^t-\sigma_H^b)\bm{E}=-d_0\sigma_H^d\bm{E},\\
    \bm{\mathcal{M}}^q & =-d_0^2(\sigma_H^t+\sigma_H^b)\bm{E}=-\frac{1}{3}d_0^2\sigma_H^q\bm{E},
\end{align}
where $d_0$ is the spacing between adjacent layers. One observes that the induced in-plane longitudinal magnetic dipole and quadrupole are compactly expressed by the electric dipole and quadrupole Hall conductivities, respectively. 

{For a semi-quantitative estimation, we take electric field strength $E=10^6$~V/m, interlayer spacing $d_0 = 3.35$~\AA, and Hall conductivity $\sigma_H^d\sim\sigma_H^q\sim 10e^2/h$, then the in-plane magnetic dipole and quadrupole can be as large as $\mathcal{M}^d\sim10^{-2}\mu_B/\text{nm}^2$ and $\mathcal{M}^q\sim10^{-3}\mu_B/\text{nm}$ respectively, where $\mu_B$ is the Bohr magneton.}

\section{Electric quadrupole Hall effect in symmetrically twisted trilayer graphene}

Symmetrically twisted trilayers are ideal platforms for studying the pure electric quadrupole Hall effect, as the dipole Hall effect is suppressed by symmetry. The interlayer translation and twist angles between adjacent layers also allow one to tune the magnitude of the currents.

\begin{figure}[t]
    \centering
    \includegraphics[width=\linewidth]{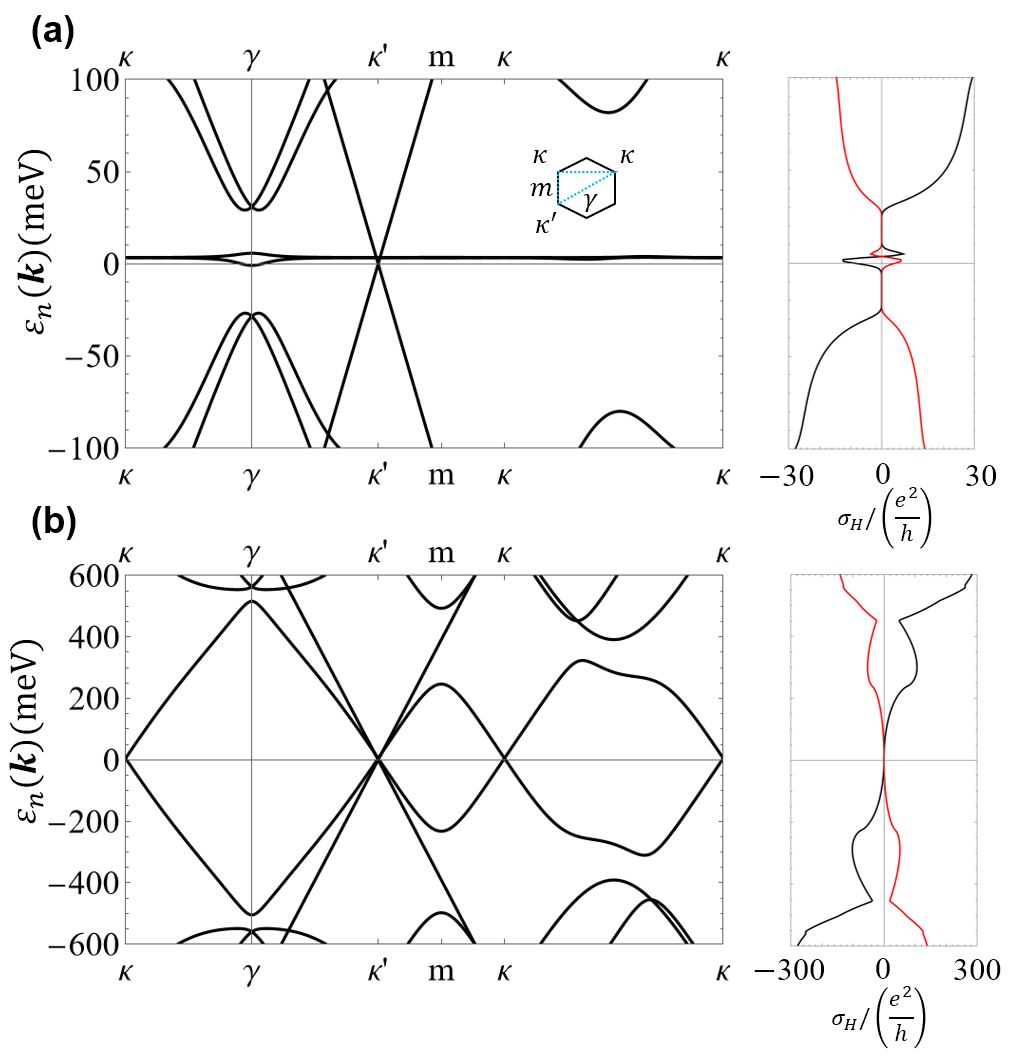}
    \caption{The moir\'e bands (left panel) and layer Hall conductivities (right panel) of mirror-symmetric TTG with twist angle $1.5^\circ$ (a) and $5^\circ$ (b). In the conductivity plots, the red curves are for the outer layers and the black curves are for the middle layer. The inset in (a) illustrates the Brillouin zone and the high-symmetry points. $\tau=1$~ps throughout this work.}\label{fig2}
\end{figure}

\begin{figure*}[t]
    \centering
    \includegraphics[width=\linewidth]{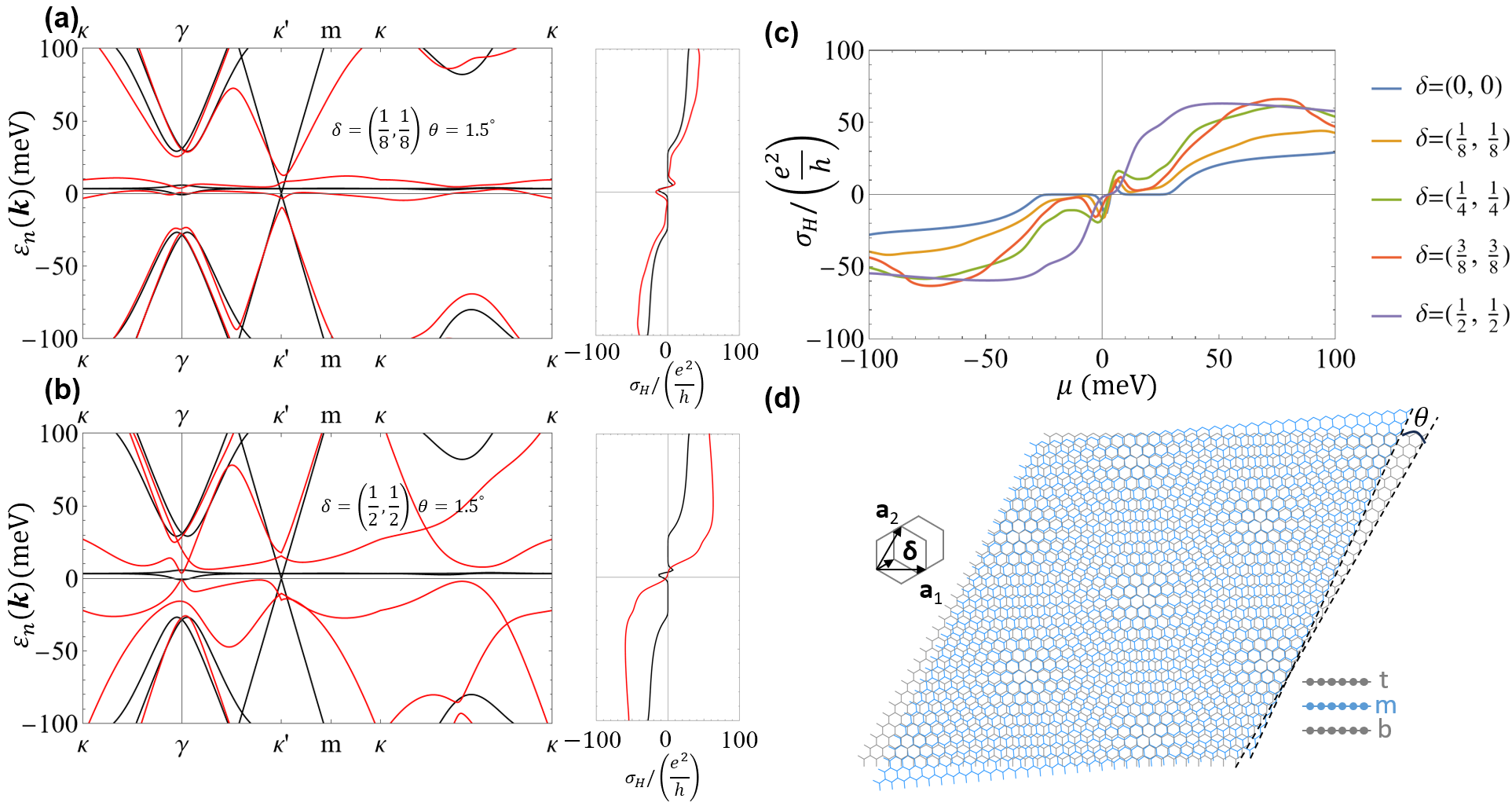}
    \caption{The moir\'e bands (left panel) and middle-layer Hall conductivity (right panel) of top-layer-translated TTG with translation \(\eta_1=\eta_2=\) (a) \(1/8\) (b) \(1/2\). The black curves are for zero translation and the red curves are for non-zero translations. (c) The middle-layer Hall conductivity of different translations and chemical potentials. The twist angle is fixed at $1.5^\circ$ in (a--c). (d) {Schematics of a mirror-symmetric twisted trilayer, where the top/bottom layers (gray) are aligned. The left inset illustrates the top-layer translation with \(\bm{\delta}=\eta_1\bm{a}_1+\eta_2\bm{a}_2\).}}\label{fig3}
\end{figure*}

\begin{figure*}[t]
    \centering
    \includegraphics[width=\linewidth]{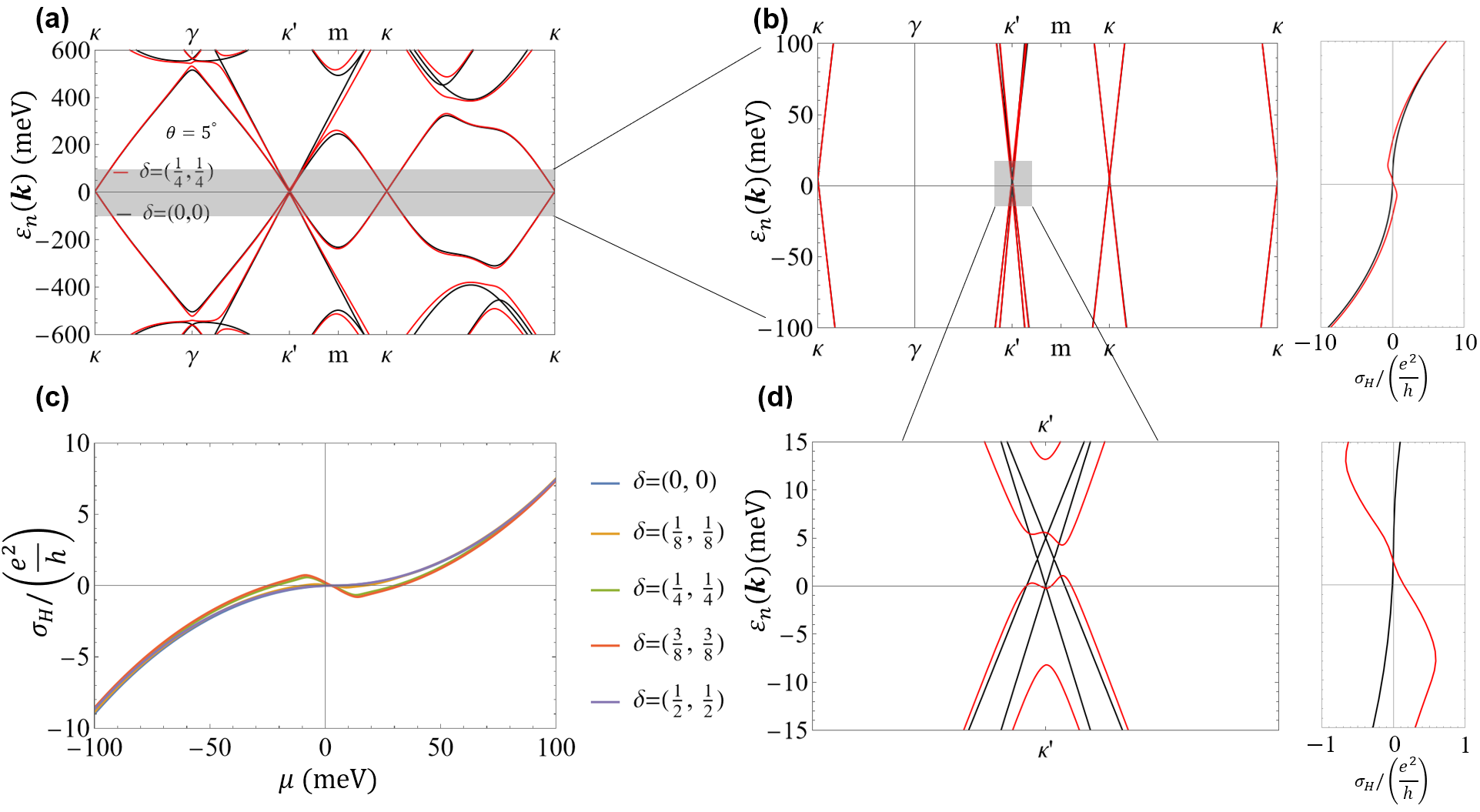}
    \caption{(a) The moir\'e bands of top-layer-translated TTG with translation \(\delta=(1/4,1/4)\) and twist angle $5^\circ$. The black curves are for zero translation and the red curves are for non-zero translation. (b, d) Zoom-in moir\'e bands (left panel) and middle-layer Hall conductivity (right panel) in the energy window of $[-100,100]$~meV and $[-15,15]$~meV, respectively. (c) The middle-layer Hall conductivity of top-layer translated TTG with different translations for $5^\circ$ twist in the energy window of $[-100,100]$~meV.}\label{fig4}
\end{figure*}

\begin{figure*}[t]
    \centering
    \includegraphics[width=0.8\linewidth]{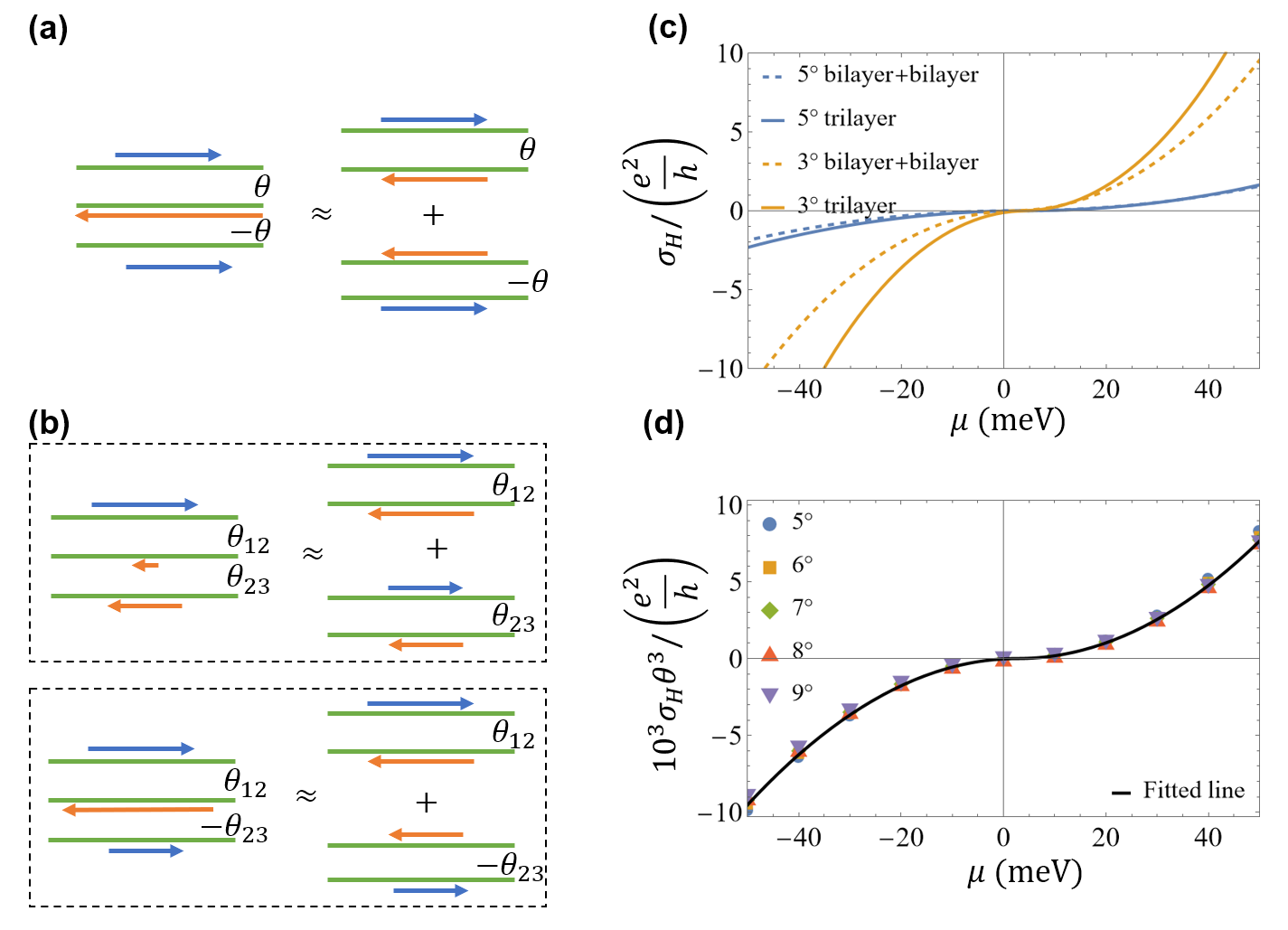}
    \caption{(a) Schematics for perturbative calculation of layer-resolved Hall currents in mirror-symmetric TTG. (b) Schematics for perturbative approximation of layer-resolved Hall currents in TTG with general twist angles: (top) helical twisted trilayer, (bottom) alternating twisted trilayer. Here $\theta_{12},\,\theta_{23}>0$. (c) The middle-layer Hall conductivities of TTG with twist angle {$3^\circ$ (orange) and $5^\circ$ (blue)}. The {solid} curves are from direct calculation of a trilayer, the {dashed} curves are calculated by adding up the contributions of two interfaces of TBG. (d) The middle-layer Hall conductivities of mirror-symmetric TTG with larger twist angles (scattered symbols), which exhibits a \(\theta^3\) scaling. The black curve is the fitted curve $f(\mu)$ (see text). The unit of \(\theta\) is rad. }\label{fig5}
\end{figure*}

\subsection{Mirror-symmetric twisted trilayer graphene}

We first consider mirror-symmetric TTG, where the outer layers are aligned and rotated with respect to the middle layer by angle $\theta$ [Fig.~\ref{fig3}(d)].
The mirror-symmetric TTG Hamiltonian in the $+K$ valley reads~\cite{TTGAshvinPRB2019,TTGNanoLett2020}
\begin{equation}
\begin{aligned}
    &H_{\rm mTTG}=\\
    &\begin{pmatrix}
        h_0(\bm{K}_t,\frac{\theta}{2}) & \mathcal{U}_{12} & 0 \\
        \mathcal{U}^{\dagger}_{12} & h_0(\bm{K}_m,-\frac{\theta}{2}) & \mathcal{U}^{\dagger}_{23} \\
        0 & \mathcal{U}_{23} & h_0(\bm{K}_b,\frac{\theta}{2})
        \end{pmatrix},
\end{aligned}
\end{equation}
where $h_0(\bm{K},\theta)=-\hbar v_F\left(-i\nabla-\bm{K}\right) \cdot R_{\theta}\left(-\sigma_x, \sigma_y\right)$ describes the twisted Dirac cone, $R_{\theta}$ denotes the rotation matrix by angle $\theta$ and $v_F=0.8 \times 10^6$~m/s is the Fermi velocity of graphene. 
The Dirac points are located at $\bm{K}_t=\bm{K}_b=R_{\frac{\theta}{2}} \bm{K}_0$ and $\bm{K}_m=R_{-\frac{\theta}{2}} \bm{K}_0$ in the three layers, where $\bm{K}_0=\left(\frac{4 \pi}{3 a}, 0\right)$ and $a=2.46$~\AA.
The interlayer coupling terms read $\mathcal{U}_{12}=\mathcal{U}_{23}=U_0+U_1 e^{-i \boldsymbol{G}_1 \cdot \boldsymbol{r}}+U_2 e^{-i\left(\boldsymbol{G}_1+\boldsymbol{G}_2\right) \cdot \boldsymbol{r}}$, where \(\bm{G}_1=\frac{2\pi}{b}(\frac{1}{\sqrt{3}},1)\), \(\bm{G}_2=\frac{2\pi}{b}(\frac{2}{\sqrt{3}},0)\), \(b=\dfrac{a}{2\sin(\theta/2)}\),
$
        U_n=\left(\begin{array}{cc}
        u_{A A} & u_{A B}e^{i \frac{2 \pi}{3}n} \\
        u_{A B}e^{-i \frac{2 \pi}{3}n} & u_{A A}
        \end{array}\right)
$,
with $u_{AA}=79.7$~meV and $u_{AB}=97.5$~meV~\cite{TBGmodel_Koshino_2018}.

Figure~\ref{fig2} shows the energy bands of $H_{\rm mTTG}$ and the corresponding layer-resolved Hall conductivities under two different twist angles. 
In the band structure, there is a Dirac cone at the $\kappa '$ point uncoupled from other bands due to the out-of-plane mirror symmetry \(M_h\)~\cite{TTGAshvinPRB2019,TTGNanoLett2020}, which can be understood as follows.
The basis of the trilayer's Bloch states can be categorized according to the eigenvalue of \(M_h\): two even states \(\ket{m}\), \(\ket{t}+\ket{b}\) and one odd state \(\ket{t}-\ket{b}\).
The $M_h$-odd state is decoupled from the even ones, retaining the linear dispersion of the monolayers and has no contribution to the layer Hall currents.
This can be explicitly shown by performing a transformation
\begin{equation}
    \mathcal{Q}=\frac{1}{\sqrt{2}}
    \begin{pmatrix}
        1 & 0 & 1\\
        0 & \sqrt{2} & 0\\
        -1 & 0 & 1
    \end{pmatrix},
\end{equation}
which changes the basis to states \(\ket{t}+\ket{b}\), \(\ket{m}\) and \(\ket{t}-\ket{b}\) via mixing \(\ket{t}\) and \(\ket{b}\) in the first and last line.
For mirror-symmetric TTG, we have \(\mathcal{U}_{12}=\mathcal{U}_{23}\) and \(\bm{K}_t=\bm{K}_b\), then the transformed Hamiltonian is block diagonal:
\begin{equation}
\begin{aligned}
&\mathcal{Q}H_{\mathrm{mTTG}}\mathcal{Q}^\dagger=\\
&\begin{pmatrix}
        h_0(\bm{K}_t,\frac{\theta}{2}) & \sqrt{2}\mathcal{U}_{12} & 0 \\
        \sqrt{2}\mathcal{U}^{\dagger}_{12} & h_0(\bm{K}_m,-\frac{\theta}{2}) & 0 \\
        0 & 0 & h_0(\bm{K}_t,\frac{\theta}{2})
    \end{pmatrix}.
\end{aligned}
\end{equation}
The top-left $2\times2$ block is equivalent to the model of TBG~\cite{BM,TBGmodel_Koshino_2018} with enlarged interlayer tunneling, which is responsible for the layer Hall currents. Note the Dirac point energy from this effective TBG model (e.g., at $\kappa$ in Fig.~\ref{fig2}) is slightly higher than zero due to interlayer coupling~\cite{TTGNanoLett2020}.
The Hall currents in the top and bottom layers are equal, ensured by the $M_h$ symmetry; the Hall current in the middle layer flows oppositely with doubled magnitude, ensuring a null net Hall current from the trilayer by the Onsager relation~\cite{TobiasTrilayerPRB2024}. 
The Hall conductivity of the layer can be very large (e.g. $10\sim100\,e^2/h$ with a relaxation time of $\tau=1$~ps) and its profile is similar to that of TBG~\cite{TobiasPRL2018,TobiasPRB2018,TobiasPRB2020,ZhaiLayerHallNC2023,ZhuJihangLayerHall2024} as expected, which is nearly antisymmetric along the energy axis and peaks at energies with strong interlayer coupling.
Overall, the Hall currents in the three layers constitute an interlayer electric quadrupole Hall current, and the resultant opposite in-plane magnetic dipole moments contributed by the two twisted interfaces lead to an in-plane magnetic quadrupole moment (Fig.~\ref{fig1}).

\subsection{Effects of interlayer translation}

Keeping the twisted angle unchanged, the electronic properties of TTG can be tuned by the lateral shift between the two outer layers.
Starting from a mirror-symmetric TTG, we assume that the top layer is translated by  $\bm{\delta}$ with respect to the bottom layer [Fig.~\ref{fig3}(d)].
The translation enters the Hamiltonian through the interlayer coupling between the top and middle layers.
To see this, it is convenient to rewrite the interlayer tunneling term as $\mathcal{U}_{12}=U_0+U_1 e^{-i \bm{b}_1 \cdot \bm{d}}+U_2 e^{-i\left(\bm{b}_1+\bm{b}_2\right) \cdot \bm{d}}$, where $\bm{d}=\left(R_{\frac{\theta}{2}}-R_{-\frac{\theta}{2}}\right) \bm{r}$ is the local interlayer displacement vector due to rotation.
When the top layer is translated by \(\bm{\delta}\), the local interlayer displacement vector becomes \(\bm{d}+\bm{\delta}\), thus the interlayer coupling between the top and middle layers becomes $\tilde{\mathcal{U}}_{12} = U_0+U_1 e^{-i \bm{b}_1 \cdot \bm{\delta}} e^{-i \bm{G}_1 \cdot \bm{r}}+U_2 e^{-i\left(\bm{b}_1+\bm{b}_2\right) \cdot \bm{\delta}} e^{-i\left(\bm{G}_1+\bm{G}_2\right) \cdot \bm{r}}$.
The effects of top layer translation are thus encoded in the two extra phases.
In the following, we parameterize \(\bm{\delta}=(\eta_1,\eta_2)=\eta_1 \bm{a}_1 + \eta_2 \bm{a}_2\), where $\eta_{1,2}\in[0,1]$ are some constants [Fig.~\ref{fig3}(d)].
Note that although the \(M_h\) symmetry is broken by the translation, the \(C_2M_h\) symmetry is respected~\cite{TTGMacDonaldPRL2021}, thus the Hall currents in the outer layers remain identical, whose sum is canceled by the Hall current in the middle layer. Therefore, in the translated TTG, the electric dipole Hall effect is still forbidden, and the layer Hall currents manifest as a pure quadrupole Hall effect.

The red curves in Figs.~\ref{fig3}(a, b) present the energy bands and the middle-layer Hall conductivity of TTG with a small twist angle ($1.5^\circ$) and two different top-layer translations.
As the translation breaks the \(M_h\) symmetry, the $M_h$-odd state \(\ket{t}-\ket{b}\) is coupled to the even states, opening a gap in the Dirac cone at the \(\kappa '\) point.
For small twist angles, the energy bands change dramatically due to this additional state entering the interlayer coupling.
Correspondingly, the profile and magnitude of layer Hall conductivity is extremely sensitive to the value of translation [Fig.~\ref{fig3}(c)]. One notices that the layer Hall conductivity generally increases with the translation.
In contrast, for large twist angles, the effects of interlayer coupling are pronounced at much higher energies [Fig.~\ref{fig4}(a)], while the low-energy bands are barely affected by the translation [Fig.~\ref{fig4}(b)]. Nevertheless, the layer Hall conductivity is sizable at low energies and it is weakly affected by the interlayer translation, where a modest fluctuation is only observable near the charge neutrality point [Fig.~\ref{fig4}(c)] due to the small gap openings at the $\kappa'$ point [Fig.~\ref{fig4}(d)]. The weak interlayer coupling and translation-independence at low energies in large-angle TTG make it possible to extend the above discussions to general multilayers consisting of varying twist angles that do not even have a well-defined periodicity.

\section{Twisted multilayer graphene with different twist angles: A Perturbative view}

In large-angle TTG, the momentum mismatch of the low energy valleys between adjacent layers is substantially larger than the interlayer coupling energy scale.
Consequently, within the low-energy window near the charge neutrality point, the interlayer coupling acts as a weak perturbation~\cite{ZhaiLayerHallNC2023}. 
For the middle layer of a TTG system, the electronic states are subjected to two distinct moir\'e potentials originating from the top and bottom twisted interfaces. 
In this large-angle perturbative regime, higher-order scattering processes that cross-couple both interfaces can be safely ignored. 
Therefore, one can decouple the trilayer system and treat the physical quantities of the middle layer as a simple linear superposition of the independent contributions from the two adjacent TBG interfaces.

To examine this, we evaluate the middle-layer Hall conductivity of mirror-symmetric TTG in two ways: (i) by performing the calculation exactly using a full trilayer Hamiltonian; (ii) by adding the separated contributions from two individual TBG systems of opposite twist angles [Fig.~\ref{fig5}(a) orange arrows].
As shown in Fig.~\ref{fig5}(c), the Hall conductivity of the middle-layer for mirror-symmetric TTG with large twist angles $\theta \gtrsim 5^\circ$ can be well approximated by the independent contributions from two TBG interfaces, i.e. $\sigma_{H}^{m}(\theta,-\theta)\approx\tilde{\sigma}_{H}^{b}(\theta)+\tilde{\sigma}_{H}^{t}(-\theta)$, where $\theta$ and $-\theta$ denote the twist angles at the bottom and top interfaces, respectively.
Here, $\tilde{\sigma}_{H}^{l}(\theta)$ denotes the Hall conductivity in the layer $l$ of the TBG with the twist angle $\theta$, satisfying $\tilde{\sigma}_{H}^{t}(\theta)=-\tilde{\sigma}_{H}^{b}(\theta)=-\tilde{\sigma}_{H}^{t}(-\theta)$~\cite{ZhaiLayerHallNC2023}.
Furthermore, we note that the layer Hall conductivity at large twist angles exhibits a robust scaling behavior: $\sigma_{H}^{m}(\theta,-\theta)\approx2\tilde{\sigma}_{H}^{b}(\theta)\approx\theta^{-3}f(\mu)e^2/h$, where $f(\mu)=C(\mu-\mu_0)|\mu-\mu_0|$ with fitting parameters $C=3.426\times10^{-6}~\mathrm{meV}^{-2}$ and $\mu_0=2.747$~meV. 
The small offset $\mu_0$ originates from the slight shift of the charge neutrality point due to the weak interlayer coupling.

The validity of this linear additive rule in TTG provides a powerful framework for generalizing to more complex multilayer geometries. 
As long as all adjacent twist angles remain in the large-angle perturbative regime, the absence of cross-interface coupling allows us to completely decompose the multilayer system into a series of decoupled TBG interfaces.
This feature naturally extends our analysis to multilayer systems containing diverse twist angle combinations, explicitly bypassing the computational difficulty of multilayer with arbitrary twist angle combination comes from the absence of long range periodicity. 
As an example of the simplest multilayer, Fig.~\ref{fig5}(b) schematically illustrates helical TTG (top)~\cite{ZhaoPeiTwistedhBN2021, HelicalTTGTrithepSciAdv2023, HelicalTTGPabloNatPhys2025} and alternating TTG (bottom), demonstrating how the total layer-resolved response can be straightforwardly assembled from the relevant isolated twisted bilayer contributions.

\section{Discussion and Summary}

In this work, we focus on exploring the effects of twist angle and interlayer translation on the layer-dependent Hall currents in TTG. 
In alternating TTG with opposite twist angles $\pm\theta$ at the two twisted interfaces, we show that the Hall currents in the top and bottom layers are identical, while the middle layer holds an opposite Hall current of doubled amplitude (Fig.~\ref{fig1}). 
This specific layer-dependent Hall current configuration also corresponds to a pure interlayer electric quadrupole Hall current and generates an in-plane magnetic quadrupole moment.
When $\theta$ is small, interlayer translation can affect the layer Hall currents dramatically, serving as a tuning knob for controlling the magnitude of the currents. 
When $\theta$ is large, interlayer translation has minor effects on the magnitude of the layer Hall currents at low energies. Importantly, in this large-angle regime the layer Hall currents in the TTG can be accurately reproduced by adding the contributions from the two individual twisted interfaces [Fig.~\ref{fig5}(a)]. 
We also identify a universal scaling behavior of the layer Hall current contributed by large-angle twisted interfaces with respect to twist angle $\theta$ and Fermi energy. 
This makes it possible to obtain the layer Hall currents in generic multilayers with twisted interfaces of varying angles by decomposing the multilayer stack into a series of individual bilayer interfaces and summing over their layer Hall currents [Fig.~\ref{fig5}(b)]. 
The layer Hall currents can be further tuned by applying an interlayer displacement field: Due to the breaking of $M_h$ and $C_2M_h$ symmetries, the Hall currents in the top and bottom layers will be different, pointing to the presence of an interlayer electric dipole Hall effect and in-plane magnetic dipole moment in addition to their quadrupole counterparts {(see Sec.~S1 in the Supporting Information for details)}.
{Experimentally, the layer-dependent Hall currents can be probed with layer-resolved measurements~\cite{ZhaiLayerHallNC2023,Cao2026}, the details of which are discussed in Sec.~S2 of the Supporting Information.}
Our work shows that twisted van der Waals materials are promising platforms for exploring layertronics responses with rich internal structures. {Our findings could stimulate future studies on understanding fundamental magnetic properties of quasi-2D systems~\cite{HuiyuanPlanarHall,LayerHallFanFengrenNC2024,HuJinxinInplaneOrbital} and seeking potential applications of twisted materials for magnetoelecctric effects~\cite{MagnetoElectricEffect2019,MagnetoElectricEffect2005}.}

\begin{acknowledgement}
The work is supported by the National Natural Science Foundation of China (No. 12425406), Research Grant Council of Hong Kong (AoE/P-701/20, HKU SRFS21227S05), and New Cornerstone Science Foundation. C.X. is sponsored by National Natural Science Foundation of China (Grant No. 12574114) and the start-up funding from Fudan University.
\end{acknowledgement}

\bibliography{references}

@article{ZhaiLayerHallNC2023,
author={Zhai, Dawei
and Chen, Cong
and Xiao, Cong
and Yao, Wang},
title={Time-reversal even charge hall effect from twisted interface coupling},
journal={Nat. Commun.},
year={2023},
month={Apr},
day={07},
volume={14},
number={1},
pages={1961},
issn={2041-1723},
doi={10.1038/s41467-023-37644-0},
url={https://doi.org/10.1038/s41467-023-37644-0}
}

@article{BM,
author = {Rafi Bistritzer  and Allan H. MacDonald },
title = {Moiré bands in twisted double-layer graphene},
journal = {Proc. Natl. Acad. Sci. U.S.A.},
volume = {108},
number = {30},
pages = {12233-12237},
year = {2011},
doi = {10.1073/pnas.1108174108},
URL = {https://www.pnas.org/doi/abs/10.1073/pnas.1108174108}
}

@article{HuiyuanMultipole2024,
author = {Zheng, Huiyuan and Zhai, Dawei and Xiao, Cong and Yao, Wang},
title = {Interlayer Electric Multipoles Induced by In-Plane Field from Quantum Geometric Origins},
journal = {Nano Lett.},
volume = {24},
number = {26},
pages = {8017-8023},
year = {2024},
doi = {10.1021/acs.nanolett.4c01657},
URL = {https://doi.org/10.1021/acs.nanolett.4c01657}
}

@article{ChenCongPRR2024,
  title = {Crossed nonlinear dynamical Hall effect in twisted bilayers},
  author = {Chen, Cong and Zhai, Dawei and Xiao, Cong and Yao, Wang},
  journal = {Phys. Rev. Res.},
  volume = {6},
  issue = {1},
  pages = {L012059},
  numpages = {6},
  year = {2024},
  month = {Mar},
  publisher = {American Physical Society},
  doi = {10.1103/PhysRevResearch.6.L012059},
  url = {https://link.aps.org/doi/10.1103/PhysRevResearch.6.L012059}
}

@Article{LiJuncheng2024,
author={Li, Juncheng
and Zhai, Dawei
and Xiao, Cong
and Yao, Wang},
title={Dynamical chiral Nernst effect in twisted Van der Waals few layers},
journal={Quantum Front.},
year={2024},
month={Jun},
day={07},
volume={3},
number={1},
pages={11},
issn={2731-6106},
doi={10.1007/s44214-024-00059-z},
url={https://doi.org/10.1007/s44214-024-00059-z}
}

@article{ZhuJihangLayerHall2024,
  title = {Layer Hall counterflow as a model probe of magic-angle twisted bilayer graphene},
  author = {Zhu, Jihang and Zhai, Dawei and Xiao, Cong and Yao, Wang},
  journal = {Phys. Rev. B},
  volume = {109},
  issue = {15},
  pages = {155114},
  numpages = {13},
  year = {2024},
  month = {Apr},
  publisher = {American Physical Society},
  doi = {10.1103/PhysRevB.109.155114},
  url = {https://link.aps.org/doi/10.1103/PhysRevB.109.155114}
}

@article{YaoYuguiLayerNernstPRB2024,
  title = {Colossal layer Nernst effect in twisted moir\'e layers},
  author = {Hu, Jin-Xin and Zeng, Chuanchang and Yao, Yugui},
  journal = {Phys. Rev. B},
  volume = {109},
  issue = {20},
  pages = {L201403},
  numpages = {6},
  year = {2024},
  month = {May},
  publisher = {American Physical Society},
  doi = {10.1103/PhysRevB.109.L201403},
  url = {https://link.aps.org/doi/10.1103/PhysRevB.109.L201403}
}

@Article{TBGYangBinghaiInnovation2021,
author={Liu, Yizhou
and Holder, Tobias
and Yan, Binghai},
title={Chirality-Induced Giant Unidirectional Magnetoresistance in Twisted Bilayer Graphene},
journal={Innovation},
year={2021},
month={Feb},
day={28},
publisher={Elsevier},
volume={2},
number={1},
issn={2666-6758},
doi={10.1016/j.xinn.2021.100085},
pages = {100085},
url={https://doi.org/10.1016/j.xinn.2021.100085}
}

@article{TobiasPRL2018,
  title = {Chiral Response of Twisted Bilayer Graphene},
  author = {Stauber, T. and Low, T. and G\'omez-Santos, G.},
  journal = {Phys. Rev. Lett.},
  volume = {120},
  issue = {4},
  pages = {046801},
  numpages = {6},
  year = {2018},
  month = {Jan},
  publisher = {American Physical Society},
  doi = {10.1103/PhysRevLett.120.046801},
  url = {https://link.aps.org/doi/10.1103/PhysRevLett.120.046801}
}

@article{TobiasPRB2018,
  title = {Linear response of twisted bilayer graphene: Continuum versus tight-binding models},
  author = {Stauber, T. and Low, T. and G\'omez-Santos, G.},
  journal = {Phys. Rev. B},
  volume = {98},
  issue = {19},
  pages = {195414},
  numpages = {11},
  year = {2018},
  month = {Nov},
  publisher = {American Physical Society},
  doi = {10.1103/PhysRevB.98.195414},
  url = {https://link.aps.org/doi/10.1103/PhysRevB.98.195414}
}

@article{TobiasPRB2020,
  title = {Change of chirality at magic angles of twisted bilayer graphene},
  author = {Stauber, T. and Gonz\'alez, J. and G\'omez-Santos, G.},
  journal = {Phys. Rev. B},
  volume = {102},
  issue = {8},
  pages = {081404},
  numpages = {5},
  year = {2020},
  month = {Aug},
  publisher = {American Physical Society},
  doi = {10.1103/PhysRevB.102.081404},
  url = {https://link.aps.org/doi/10.1103/PhysRevB.102.081404}
}

@article{TobiasNanoscale2020,
author ={Bahamon, Dario A. and Gómez-Santos, G. and Stauber, T.},
title  ={Emergent magnetic texture in driven twisted bilayer graphene},
journal  ={Nanoscale},
year  ={2020},
volume  ={12},
issue  ={28},
pages  ={15383-15392},
publisher  ={The Royal Society of Chemistry},
doi  ={10.1039/D0NR02786C},
url  ={http://dx.doi.org/10.1039/D0NR02786C}
}

@article{TobiasNanoLett2020,
author = {Stauber, Tobias and Low, Tony and Gómez-Santos, Guillermo},
title = {Plasmon-Enhanced Near-Field Chirality in Twisted van der Waals Heterostructures},
journal = {Nano Lett.},
volume = {20},
number = {12},
pages = {8711-8718},
year = {2020},
doi = {10.1021/acs.nanolett.0c03519},
URL = {https://doi.org/10.1021/acs.nanolett.0c03519}
}

@article{TobiasTrilayerPRB2024,
  title = {Optical response of alternating twisted trilayer graphene},
  author = {Margetis, Dionisios and G\'omez-Santos, Guillermo and Stauber, Tobias},
  journal = {Phys. Rev. B},
  volume = {110},
  issue = {20},
  pages = {205144},
  numpages = {17},
  year = {2024},
  month = {Nov},
  publisher = {American Physical Society},
  doi = {10.1103/PhysRevB.110.205144},
  url = {https://link.aps.org/doi/10.1103/PhysRevB.110.205144}
}

@article{Brey2DM2017,
doi = {10.1088/2053-1583/aa7eb6},
url = {https://dx.doi.org/10.1088/2053-1583/aa7eb6},
year = {2017},
month = {jul},
publisher = {IOP Publishing},
volume = {4},
number = {3},
pages = {035015},
author = {E Suárez Morell and Leonor Chico and Luis Brey},
title = {Twisting dirac fermions: circular dichroism in bilayer graphene},
journal = {2D Mater.}
}

@Article{CDTBGNatNano2016,
author={Kim, Cheol-Joo
and S{\'a}nchez-Castillo, A.
and Ziegler, Zack
and Ogawa, Yui
and Noguez, Cecilia
and Park, Jiwoong},
title={Chiral atomically thin films},
journal={Nat. Nanotechnol.},
year={2016},
month={Jun},
day={01},
volume={11},
number={6},
pages={520-524},
issn={1748-3395},
doi={10.1038/nnano.2016.3},
url={https://doi.org/10.1038/nnano.2016.3}
}

@article{CDTwistedhBNPRL2020,
  title = {Flat Bands and Chiral Optical Response of Moir\'e Insulators},
  author = {Ochoa, H. and Asenjo-Garcia, A.},
  journal = {Phys. Rev. Lett.},
  volume = {125},
  issue = {3},
  pages = {037402},
  numpages = {5},
  year = {2020},
  month = {Jul},
  publisher = {American Physical Society},
  doi = {10.1103/PhysRevLett.125.037402},
  url = {https://link.aps.org/doi/10.1103/PhysRevLett.125.037402}
}

@article{CDTBGslidingMelePRB2019,
  title = {Twist, slip, and circular dichroism in bilayer graphene},
  author = {Addison, Zachariah and Park, Jiwoong and Mele, E. J.},
  journal = {Phys. Rev. B},
  volume = {100},
  issue = {12},
  pages = {125418},
  numpages = {7},
  year = {2019},
  month = {Sep},
  publisher = {American Physical Society},
  doi = {10.1103/PhysRevB.100.125418},
  url = {https://link.aps.org/doi/10.1103/PhysRevB.100.125418}
}

@article{LiCiChiralExcitonHall2024,
  title = {Chiral excitonic systems in twisted bilayers from F\"orster coupling and unconventional excitonic Hall effects},
  author = {Li, Ci and Yao, Wang},
  journal = {Phys. Rev. B},
  volume = {110},
  issue = {12},
  pages = {L121407},
  numpages = {8},
  year = {2024},
  month = {Sep},
  publisher = {American Physical Society},
  doi = {10.1103/PhysRevB.110.L121407},
  url = {https://link.aps.org/doi/10.1103/PhysRevB.110.L121407}
}

@article{Onsager,
  title = {Reciprocal Relations in Irreversible Processes. I.},
  author = {Onsager, Lars},
  journal = {Phys. Rev.},
  volume = {37},
  issue = {4},
  pages = {405--426},
  numpages = {0},
  year = {1931},
  month = {Feb},
  publisher = {American Physical Society},
  doi = {10.1103/PhysRev.37.405},
  url = {https://link.aps.org/doi/10.1103/PhysRev.37.405}
}

@article{TBGmodel_Koshino_2018,
  title = {Maximally Localized Wannier Orbitals and the Extended Hubbard Model for Twisted Bilayer Graphene},
  author = {Koshino, Mikito and Yuan, Noah F. Q. and Koretsune, Takashi and Ochi, Masayuki and Kuroki, Kazuhiko and Fu, Liang},
  journal = {Phys. Rev. X},
  volume = {8},
  issue = {3},
  pages = {031087},
  numpages = {12},
  year = {2018},
  month = {Sep},
  publisher = {American Physical Society},
  doi = {10.1103/PhysRevX.8.031087},
  url = {https://link.aps.org/doi/10.1103/PhysRevX.8.031087}
}

@article{TTGAshvinPRB2019,
  title = {Magic angle hierarchy in twisted graphene multilayers},
  author = {Khalaf, Eslam and Kruchkov, Alex J. and Tarnopolsky, Grigory and Vishwanath, Ashvin},
  journal = {Phys. Rev. B},
  volume = {100},
  issue = {8},
  pages = {085109},
  numpages = {9},
  year = {2019},
  month = {Aug},
  publisher = {American Physical Society},
  doi = {10.1103/PhysRevB.100.085109},
  url = {https://link.aps.org/doi/10.1103/PhysRevB.100.085109}
}

@article{TTGNanoLett2020,
author = {Carr, Stephen and Li, Chenyuan and Zhu, Ziyan and Kaxiras, Efthimios and Sachdev, Subir and Kruchkov, Alexander},
title = {Ultraheavy and Ultrarelativistic Dirac Quasiparticles in Sandwiched Graphenes},
journal = {Nano Lett.},
volume = {20},
number = {5},
pages = {3030-3038},
year = {2020},
doi = {10.1021/acs.nanolett.9b04979},
URL = { https://doi.org/10.1021/acs.nanolett.9b04979}
}

@article{TTGMacDonaldPRL2021,
  title = {In-Plane Critical Magnetic Fields in Magic-Angle Twisted Trilayer Graphene},
  author = {Qin, Wei and MacDonald, Allan H.},
  journal = {Phys. Rev. Lett.},
  volume = {127},
  issue = {9},
  pages = {097001},
  numpages = {7},
  year = {2021},
  month = {Aug},
  publisher = {American Physical Society},
  doi = {10.1103/PhysRevLett.127.097001},
  url = {https://link.aps.org/doi/10.1103/PhysRevLett.127.097001}
}

@Article{LayerHallNature2011,
author={Gao, Anyuan
and Liu, Yu-Fei
and Hu, Chaowei
and Qiu, Jian-Xiang
and Tzschaschel, Christian
and Ghosh, Barun
and Ho, Sheng-Chin
and B{\'e}rub{\'e}, Damien
and Chen, Rui
and Sun, Haipeng
and Zhang, Zhaowei
and Zhang, Xin-Yue
and Wang, Yu-Xuan
and Wang, Naizhou
and Huang, Zumeng
and Felser, Claudia
and Agarwal, Amit
and Ding, Thomas
and Tien, Hung-Ju
and Akey, Austin
and Gardener, Jules
and Singh, Bahadur
and Watanabe, Kenji
and Taniguchi, Takashi
and Burch, Kenneth S.
and Bell, David C.
and Zhou, Brian B.
and Gao, Weibo
and Lu, Hai-Zhou
and Bansil, Arun
and Lin, Hsin
and Chang, Tay-Rong
and Fu, Liang
and Ma, Qiong
and Ni, Ni
and Xu, Su-Yang},
title={Layer Hall effect in a 2D topological axion antiferromagnet},
journal={Nature},
year={2021},
month={Jul},
day={01},
volume={595},
number={7868},
pages={521-525},
issn={1476-4687},
doi={10.1038/s41586-021-03679-w},
url={https://doi.org/10.1038/s41586-021-03679-w}
}

@article{LayerHallNSR2022,
    author = {Chen, Rui and Sun, Hai-Peng and Gu, Mingqiang and Hua, Chun-Bo and Liu, Qihang and Lu, Hai-Zhou and Xie, X C},
    title = {Layer Hall effect induced by hidden Berry curvature in antiferromagnetic insulators},
    journal = {Natl. Sci. Rev.},
    volume = {11},
    number = {2},
    pages = {nwac140},
    year = {2022},
    month = {08},
    issn = {2095-5138},
    doi = {10.1093/nsr/nwac140},
    url = {https://doi.org/10.1093/nsr/nwac140}
}

@article{LayerHallPRB2022,
  title = {Quantum anomalous layer Hall effect in the topological magnet ${\mathrm{MnBi}}_{2}{\mathrm{Te}}_{4}$},
  author = {Dai, Wen-Bo and Li, Hailong and Xu, Dong-Hui and Chen, Chui-Zhen and Xie, X. C.},
  journal = {Phys. Rev. B},
  volume = {106},
  issue = {24},
  pages = {245425},
  numpages = {5},
  year = {2022},
  month = {Dec},
  publisher = {American Physical Society},
  doi = {10.1103/PhysRevB.106.245425},
  url = {https://link.aps.org/doi/10.1103/PhysRevB.106.245425}
}

@article{LayerHallNSR2023,
    author = {Li, Shuai and Gong, Ming and Cheng, Shuguang and Jiang, Hua and Xie, X C},
    title = {Dissipationless layertronics in axion insulator MnBi2Te4},
    journal = {Natl. Sci. Rev.},
    volume = {11},
    number = {6},
    pages = {nwad262},
    year = {2023},
    month = {10},
    issn = {2095-5138},
    doi = {10.1093/nsr/nwad262},
    url = {https://doi.org/10.1093/nsr/nwad262}
}

@article{LayerHallYaoYuguiPRB2025,
  title = {Layer Hall and layer spin Hall effects in two-dimensional altermagnets induced by spin-layer coupling},
  author = {Wang, Xiangju and Liu, Siyuan and Bai, Ling and Zhang, Run-Wu and Yao, Yugui and Feng, Wanxiang},
  journal = {Phys. Rev. B},
  volume = {112},
  issue = {13},
  pages = {134421},
  numpages = {8},
  year = {2025},
  month = {Oct},
  publisher = {American Physical Society},
  doi = {10.1103/643d-wkc1},
  url = {https://link.aps.org/doi/10.1103/643d-wkc1}
}

@article{LayerHallQiaoZhenhuaPRL2025,
  title = {Layer Hall Effect without External Electric Field},
  author = {Han, Yulei and Guo, Yunpeng and Li, Zeyu and Qiao, Zhenhua},
  journal = {Phys. Rev. Lett.},
  volume = {134},
  issue = {23},
  pages = {236206},
  numpages = {7},
  year = {2025},
  month = {Jun},
  publisher = {American Physical Society},
  doi = {10.1103/tdx6-gnxz},
  url = {https://link.aps.org/doi/10.1103/tdx6-gnxz}
}

@Article{LayHallMaYandongNC2025,
author={Du, Wenhui
and Dou, Kaiying
and Li, Xinru
and Dai, Ying
and Wang, Zeyan
and Huang, Baibiao
and Ma, Yandong},
title={Topological layer Hall effect in two-dimensional type-I multiferroic heterostructure},
journal={Nat. Commun.},
year={2025},
month={Jul},
day={03},
volume={16},
number={1},
pages={6141},
issn={2041-1723},
doi={10.1038/s41467-025-61514-6},
url={https://doi.org/10.1038/s41467-025-61514-6}
}

@article{LayerHallAdvSci2021,
author = {Xu, Haowei and Zhou, Jian and Li, Ju},
title = {Light-Induced Quantum Anomalous Hall Effect on the 2D Surfaces of 3D Topological Insulators},
journal = {Adv. Sci.},
volume = {8},
number = {17},
pages = {2101508},
doi = {https://doi.org/10.1002/advs.202101508},
url = {https://advanced.onlinelibrary.wiley.com/doi/abs/10.1002/advs.202101508},
year = {2021}
}

@Article{LayerHallFanFengrenNC2024,
author={Fan, Feng-Ren
and Xiao, Cong
and Yao, Wang},
title={Intrinsic dipole Hall effect in twisted MoTe2: magnetoelectricity and contact-free signatures of topological transitions},
journal={Nat. Commun.},
year={2024},
month={Sep},
day={12},
volume={15},
number={1},
pages={7997},
issn={2041-1723},
doi={10.1038/s41467-024-52314-5},
url={https://doi.org/10.1038/s41467-024-52314-5}
}

@Article{TBGCaoYuan2018a,
author={Cao, Yuan
and Fatemi, Valla
and Fang, Shiang
and Watanabe, Kenji
and Taniguchi, Takashi
and Kaxiras, Efthimios
and Jarillo-Herrero, Pablo},
title={Unconventional superconductivity in magic-angle graphene superlattices},
journal={Nature},
year={2018},
month={Apr},
day={01},
volume={556},
number={7699},
pages={43-50},
issn={1476-4687},
doi={10.1038/nature26160},
url={https://doi.org/10.1038/nature26160}
}

@Article{FCIMoTe2Park2023,
author={Park, Heonjoon
and Cai, Jiaqi
and Anderson, Eric
and Zhang, Yinong
and Zhu, Jiayi
and Liu, Xiaoyu
and Wang, Chong
and Holtzmann, William
and Hu, Chaowei
and Liu, Zhaoyu
and Taniguchi, Takashi
and Watanabe, Kenji
and Chu, Jiun-Haw
and Cao, Ting
and Fu, Liang
and Yao, Wang
and Chang, Cui-Zu
and Cobden, David
and Xiao, Di
and Xu, Xiaodong},
title={Observation of fractionally quantized anomalous Hall effect},
journal={Nature},
year={2023},
month={Oct},
day={01},
volume={622},
number={7981},
pages={74-79},
issn={1476-4687},
doi={10.1038/s41586-023-06536-0},
url={https://doi.org/10.1038/s41586-023-06536-0}
}

@Article{FCIMoTe2Jiaqi2023,
author={Cai, Jiaqi
and Anderson, Eric
and Wang, Chong
and Zhang, Xiaowei
and Liu, Xiaoyu
and Holtzmann, William
and Zhang, Yinong
and Fan, Fengren
and Taniguchi, Takashi
and Watanabe, Kenji
and Ran, Ying
and Cao, Ting
and Fu, Liang
and Xiao, Di
and Yao, Wang
and Xu, Xiaodong},
title={Signatures of fractional quantum anomalous Hall states in twisted MoTe2},
journal={Nature},
year={2023},
month={Oct},
day={01},
volume={622},
number={7981},
pages={63-68},
issn={1476-4687},
doi={10.1038/s41586-023-06289-w},
url={https://doi.org/10.1038/s41586-023-06289-w}
}

@Article{FCIMoTe2ShanJie2023,
author={Zeng, Yihang
and Xia, Zhengchao
and Kang, Kaifei
and Zhu, Jiacheng
and Kn{\"u}ppel, Patrick
and Vaswani, Chirag
and Watanabe, Kenji
and Taniguchi, Takashi
and Mak, Kin Fai
and Shan, Jie},
title={Thermodynamic evidence of fractional Chern insulator in moir{\'e} MoTe2},
journal={Nature},
year={2023},
month={Oct},
day={01},
volume={622},
number={7981},
pages={69-73},
issn={1476-4687},
doi={10.1038/s41586-023-06452-3},
url={https://doi.org/10.1038/s41586-023-06452-3}
}

@article{FCIMoTe2PRX2023,
  title = {Observation of Integer and Fractional Quantum Anomalous Hall Effects in Twisted Bilayer ${\mathrm{MoTe}}_{2}$},
  author = {Xu, Fan and Sun, Zheng and Jia, Tongtong and Liu, Chang and Xu, Cheng and Li, Chushan and Gu, Yu and Watanabe, Kenji and Taniguchi, Takashi and Tong, Bingbing and Jia, Jinfeng and Shi, Zhiwen and Jiang, Shengwei and Zhang, Yang and Liu, Xiaoxue and Li, Tingxin},
  journal = {Phys. Rev. X},
  volume = {13},
  issue = {3},
  pages = {031037},
  numpages = {12},
  year = {2023},
  month = {Sep},
  publisher = {American Physical Society},
  doi = {10.1103/PhysRevX.13.031037},
  url = {https://link.aps.org/doi/10.1103/PhysRevX.13.031037}
}

@Article{ZhaoPeiTwistedhBN2021,
author={Zhao, Pei
and Xiao, Chengxin
and Yao, Wang},
title={Universal superlattice potential for 2D materials from twisted interface inside h-BN substrate},
journal={npj 2D Mater. Appl.},
year={2021},
month={Apr},
day={12},
volume={5},
number={1},
pages={38},
issn={2397-7132},
doi={10.1038/s41699-021-00221-4},
url={https://doi.org/10.1038/s41699-021-00221-4}
}

@article{HelicalTTGTrithepSciAdv2023,
author = {Trithep Devakul  and Patrick J. Ledwith  and Li-Qiao Xia  and Aviram Uri  and Sergio C. de la Barrera  and Pablo Jarillo-Herrero  and Liang Fu },
title = {Magic-angle helical trilayer graphene},
journal = {Sci. Adv.},
volume = {9},
number = {36},
pages = {eadi6063},
year = {2023},
doi = {10.1126/sciadv.adi6063},
URL = {https://www.science.org/doi/abs/10.1126/sciadv.adi6063}
}

@Article{HelicalTTGPabloNatPhys2025,
author={Xia, Li-Qiao
and de la Barrera, Sergio C.
and Uri, Aviram
and Sharpe, Aaron
and Kwan, Yves H.
and Zhu, Ziyan
and Watanabe, Kenji
and Taniguchi, Takashi
and Goldhaber-Gordon, David
and Fu, Liang
and Devakul, Trithep
and Jarillo-Herrero, Pablo},
title={Topological bands and correlated states in helical trilayer graphene},
journal={Nat. Phys.},
year={2025},
month={Feb},
day={01},
volume={21},
number={2},
pages={239-244},
issn={1745-2481},
doi={10.1038/s41567-024-02731-6},
url={https://doi.org/10.1038/s41567-024-02731-6}
}

@Article{MATTGPabloNature2021,
author={Park, Jeong Min
and Cao, Yuan
and Watanabe, Kenji
and Taniguchi, Takashi
and Jarillo-Herrero, Pablo},
title={Tunable strongly coupled superconductivity in magic-angle twisted trilayer graphene},
journal={Nature},
year={2021},
month={Feb},
day={01},
volume={590},
number={7845},
pages={249-255},
issn={1476-4687},
doi={10.1038/s41586-021-03192-0},
url={https://doi.org/10.1038/s41586-021-03192-0}
}

@article{MATTGPhilipKimScience2021,
author = {Zeyu Hao  and A. M. Zimmerman  and Patrick Ledwith  and Eslam Khalaf  and Danial Haie Najafabadi  and Kenji Watanabe  and Takashi Taniguchi  and Ashvin Vishwanath  and Philip Kim },
title = {Electric field–tunable superconductivity in alternating-twist magic-angle trilayer graphene},
journal = {Science},
volume = {371},
number = {6534},
pages = {1133-1138},
year = {2021},
doi = {10.1126/science.abg0399},
URL = {https://www.science.org/doi/abs/10.1126/science.abg0399}
}

@Article{Okyay2025,
author={Okyay, Mahmut Sait
and Choi, Min
and Xu, Qiang
and Di{\'e}guez, Adri{\'a}n Perez
and Del Ben, Mauro
and Ibrahim, Khaled Z.
and Wong, Bryan M.},
title={Unconventional nonlinear Hall effects in twisted multilayer 2D materials},
journal={npj 2D Mater. Appl.},
year={2025},
month={Jan},
day={07},
volume={9},
number={1},
pages={1},
issn={2397-7132},
doi={10.1038/s41699-024-00520-6},
url={https://doi.org/10.1038/s41699-024-00520-6}
}

@Article{Okyay2022,
author={Okyay, Mahmut Sait
and Sato, Shunsuke A.
and Kim, Kun Woo
and Yan, Binghai
and Jin, Hosub
and Park, Noejung},
title={Second harmonic Hall responses of insulators as a probe of Berry curvature dipole},
journal={Commun. Phys.},
year={2022},
month={Nov},
day={26},
volume={5},
number={1},
pages={303},
issn={2399-3650},
doi={10.1038/s42005-022-01086-9},
url={https://doi.org/10.1038/s42005-022-01086-9}
}

@article{doi:10.1126/science.abd5146,
author = {Amilcar Bedoya-Pinto  and Jing-Rong Ji  and Avanindra K. Pandeya  and Pierluigi Gargiani  and Manuel Valvidares  and Paolo Sessi  and James M. Taylor  and Florin Radu  and Kai Chang  and Stuart S. P. Parkin },
title = {Intrinsic 2D-XY ferromagnetism in a van der Waals monolayer},
journal = {Science},
volume = {374},
number = {6567},
pages = {616-620},
year = {2021},
doi = {10.1126/science.abd5146},
URL = {https://www.science.org/doi/abs/10.1126/science.abd5146},
eprint = {https://www.science.org/doi/pdf/10.1126/science.abd5146}}

@Article{Cao2026,
author={Cao, Leonard W.
and Wu, Chen
and Lyu, Lingyuan
and Cohen, Liam
and Samuelson, Noah
and Yan, Ziying
and Pancholi, Sneh
and Watanabe, Kenji
and Taniguchi, Takashi
and Parker, Daniel E.
and Young, Andrea F.
and Allen, Monica T.},
title={Layer-Resolved Microwave Imaging of a van der Waals Heterostructure},
journal={Nano Letters},
year={2026},
month={May},
day={20},
publisher={American Chemical Society},
volume={26},
number={19},
pages={6287-6295},
issn={1530-6984},
doi={10.1021/acs.nanolett.5c06452},
url={https://doi.org/10.1021/acs.nanolett.5c06452}
}

@article{hchg-7bq9,
  title = {X-ray magnetic circular dichroism originating from the anisotropic magnetic dipole operator in collinear altermagnets under trigonal crystal field},
  author = {Sasabe, Norimasa and Ishii, Yuta and Yamasaki, Yuichi},
  journal = {Phys. Rev. B},
  volume = {112},
  issue = {22},
  pages = {224401},
  numpages = {10},
  year = {2025},
  month = {Dec},
  publisher = {American Physical Society},
  doi = {10.1103/hchg-7bq9},
  url = {https://link.aps.org/doi/10.1103/hchg-7bq9}
}

@article{HuiyuanPlanarHall,
author = {Zheng, Huiyuan and Zhai, Dawei and Xiao, Cong and Yao, Wang},
title = {Layer Coherence Origin of Planar Hall Effect: From Charge to Multipole and Valley},
journal = {Nano Lett.},
volume = {25},
number = {25},
pages = {10096-10101},
year = {2025},
doi = {10.1021/acs.nanolett.5c01944},
URL = {https://doi.org/10.1021/acs.nanolett.5c01944}
}

@article{HuJinxinInplaneOrbital,
  title={Theory of In-Plane Orbital Magnetization with Layer Hybridization},
  author={Hu, Jin-Xin and Sun, Zi-Ting and Yao, Yugui},
  journal={arXiv:2606.18891},
  year={2026}
}

@Article{MagnetoElectricEffect2019,
author={Spaldin, N. A.
and Ramesh, R.},
title={Advances in magnetoelectric multiferroics},
journal={Nat. Mater.},
year={2019},
month={Mar},
day={01},
volume={18},
number={3},
pages={203-212},
issn={1476-4660},
doi={10.1038/s41563-018-0275-2},
url={https://doi.org/10.1038/s41563-018-0275-2}
}

@article{MagnetoElectricEffect2005,
doi = {10.1088/0022-3727/38/8/R01},
url = {https://doi.org/10.1088/0022-3727/38/8/R01},
year = {2005},
month = {apr},
publisher = {},
volume = {38},
number = {8},
pages = {R123},
author = {Fiebig, Manfred},
title = {Revival of the magnetoelectric effect},
journal = {J. Phys. D: Appl. Phys.}
}

\end{document}